\documentclass[useAMS,usenatbib]{mn2e}
\usepackage{amssymb,amsmath}
\usepackage{graphicx}
\usepackage{epstopdf}
\usepackage{float}
\usepackage[dvipsnames]{color}
\usepackage{url}

\def \apj {ApJ\ }
\def \apjl {ApJL\ }
\def \mnras {MNRAS\ }
\def \aap {A\&A\ }

\title[Selection biases in gamma ray burst afterglow correlations]{Selection biases in the gamma ray burst E$_{\rm iso}$ -- L$_{\rm opt,X}$ correlation}
\vspace{-5mm}
\author[]{D.M. Coward$^{1,2}$\thanks{E-mail:David.Coward@uwa.edu.au}, E.J. Howell$^{1}$, L. Wan$^3$, D. Macpherson$^{1,4}$ \\
$^{1}$School of Physics, University of Western Australia, Crawley WA 6009, Australia\\
$^{2}$Australian Research Council Future Fellow\\
$^{3}$School of Astronomy and Space Science, Nanjing University, Nanjing 210093, China.\\
$^{4}$ICRAR, University of Western Austalia, Crawley WA 6009, Australia\\
}

\begin{document}
\vspace{-5mm}

\pagerange{\pageref{firstpage}--\pageref{lastpage}} \pubyear{3002}

\maketitle

\label{firstpage}
\vspace{-5mm}
\begin{abstract}
Gamma ray burst (GRB) optical and X-ray afterglow luminosity is expected to correlate with the GRB isotropic equivalent kinetic energy of the outflow in the standard synchrotron model for GRB afterglows. Previous studies, using prompt GRB isotropic equivalent energy ($E_{\rm iso}$) as a proxy for isotropic equivalent kinetic energy, have generally confirmed a correlation between X-ray and optical afterglow luminosities. Assuming that GRB afterglow luminosity does not evolve strongly with redshift, we identify a strong Malmquist bias in GRB optical and X-ray afterglow luminosity data. We show that selection effects dominate the observed E$_{\rm iso}$ -- L$_{\rm opt,X}$ correlations, and have likely been underestimated in other studies. The bias is strongest for a subset of optically faint bursts $m>24$ at 24 hr with $z>2$. After removing this optical selection bias, the E$_{\rm iso}$ -- L$_{\rm opt,X}$ correlation for long GRBs is not statistically significant, but combining both long and short GRB luminosity data the correlation is significant. Using the median of the $E_{\rm iso}$ and $L_{\rm opt,X}$ distributions,
we apply the synchrotron model assuming the same power law index for short and long GRBs, but different microphysical parameter distributions. Comparing the ratio of optical and X-ray luminosities, we find tentative evidence that the fraction of post-shock energy in magnetic fields, $\epsilon_B$, could be systematically higher in SGRBs compared to LGRBs.
\end{abstract}


\section{Introduction}
Gamma Ray Bursts (GRBs) are the most energetic transients observed at cosmological distances. They have been categorised into two classes. The first class, `long', hereafter LGRB ($T_{90}>2$ s) \footnote{$T_{90}$ is the duration in which the cumulative counts are from 5\% to 95\% above background.} are linked to the core collapse of massive stars (collapsars) \citep{woos93,paczy98,mac99}. For several cases, the GRBs are firmly associated with Type Ib/c
supernovae \citep[e.g.][]{hjorth03,stan03}, suggesting they are linked to the end of massive stellar evolution. In contrast short GRBs ($T_{90}<2$ s hereafter SGRB) have a less certain origin.

The first breakthrough to understand the origin of SGRBs occurred in 2005 after the launch of the NASA {\it Swift} satellite \citep{geh04}. Prompt localizations and deep afterglow searches
yielded the first redshifts and investigations of their progenitor environments based on their host galaxies.
By 2014, about three dozen SGRBs had been localized by {\it Swift} and about 50\% have optical detections with redshift determinations.

Binary neutron star mergers (NS-NS) or neutron star--black hole (NS-BH) mergers are the favoured progenitors for SGRBs, based on the association of some SGRBs with an older stellar population \citep[e.g.][]{lrg05,z07}, as compared to LGRBs. Further evidence for the origin of SGRBs comes from observations showing that at least some SGRBs occur far from their site of origin, a consequence of possible high velocity kicks imparted to NSs at birth.

Although the bi-modal distribution is accepted as evidence for two GRB classifications, there remains ambiguity. About 20\% of \emph{Swift} SGRBs have been detected with an extended emission lasting up to 100\hspace{1mm}s (hereafter SGRB-EE) \citep{norris_06,Perley_extendedSGRB_08} leading to suggestions that different progenitor types produce these bursts  \citep{norris_2011}. \cite{2008MNRAS.385L..10T} argue that SGRB-EE could be NS-BH mergers based on their galaxy off-sets or the birth of a rapidly rotating proto-magnetar produced via NS-NS merger or accretion-induced collapse of a white dwarf \citep{Metzger_2008,bucc11}.

The standard synchrotron model describes the relationship between afterglow luminosity and isotropic kinetic energy of the prompt emission ($E_{\rm K,iso}$). Previous studies that have investigated this correlation for LGRBs include e.g. \cite{2001ApJ...547..922F,LZ2006,Amati060614,Kaneko2007,Gehrels2008}. For SGRBs, studies include \citet{Kouveliotou2004}, who used the X-ray luminosity at 10 hours \citep[see also][]{2006MNRAS.370.1946G, FanPiran} and also \citet{Nysewander2009}, who analysed both the optical $R-$band and X-ray luminosity at 11 hours. In addition to these studies, recent investigations of correlations between afterglow luminosity and $E_{\rm K,iso}$ include \citet{2010ApJ...720.1513K,2013arXiv1311.2603B,2012MNRAS.425..506D,2013MNRAS.428..729M,2013arXiv1311.2603B}.

 \cite{2010ApJ...720.1513K} (hereafter K10) analysed how the optical flux density in the $R_C$ band at one day (in the host frame assuming $z=1$) is correlated with $E_{\rm iso}$ for a sample of LGRBs. While no tight correlation was identified, they found a general trend of increasing optical luminosity with increasing $E_{\rm iso}$. 
 K10 provided a best fit to the correlation, and find that $L_{\rm opt} \propto E_{\rm iso}^{0.36}$, which is significantly shallower than the standard synchrotron model predicts i.e. $E_{\rm K,iso}^{1.1}$.
 
 \cite{2013arXiv1311.2603B} performed a similar analysis, but included 28 SGRBs X-ray and optical afterglows. X-ray luminosities were calculated at a
fiducial rest-frame time of 11 hr in the $0.3-10$ keV band
($L_{X,11}$) as a function of the isotropic-equivalent $\gamma$-ray
energy ($E_{\rm iso}$). The study showed that the observed correlations are flatter than the theoretical expectation, similar to that identified by K10.
 
\cite{2012MNRAS.425..506D}, claim they have obtained a {\it complete} selection of {\it Swift} LGRBs by applying a high cut-off in GRB peak photon flux to minimize flux bias.
They show that the X-ray afterglow luminosity vs $E_{\rm iso}$ correlation evolves from strongest at early times to weakest at late times. Despite this high-energy flux limit, they acknowledge that their {\it complete} GRB X-ray luminosity sample is still biased, and apply a joint correlation method to account for the Malmquist bias correlation between redshift and luminosity. Fig 2 shows that a cut in $\gamma$-ray peak photon flux does not remove the redshift bias in their X-ray luminosity distribution.

Given the importance of testing the standard synchrotron model using multi-wavelength GRB data, we re-examine the $L_{\rm opt,24}-E_{\rm iso}$ correlation. Our main focus is understanding how selection biases in the GRB afterglow distribution influence the $L_{\rm opt,24}-E_{\rm iso}$ correlation. We will apply robust methods to first identify flux limited biases (Malmquist) in GRB optical and X-ray afterglow data, and secondly, remove this bias to construct a bias free selection. Finally, we compare the unbiased LGRB and SGRB optical and X-ray luminosities with the standard synchrotron model.  

\subsection{GRB afterglow energetics in the Standard Fireball Model}
The standard afterglow synchrotron model uses free parameters
that describe the relativistic shock microphysics: the fraction of
post-shock energy in the magnetic fields, $\epsilon_B$, the fraction of energy in relativistic electrons, $\epsilon_e$. These parameters follow a power-law distribution i.e. $N(\gamma)\propto\gamma^{-p}$ above
a minimum Lorentz factor, $\gamma_m$ \citep{1998ApJ...497L..17S}.


%

In the standard afterglow synchrotron model, the X-ray
band is expected to be located near or above the synchrotron cooling
frequency \citep{2002ApJ...568..820G}.  The afterglow X-ray
flux (assuming a fiducial value for $p\approx2.4$) is given by:
\begin{equation}
F_{\nu,X}\propto E_{\rm K,iso}^{(p+2)/4}\,
\epsilon_e^{p-1}\, \epsilon_B^{(p-2)/4} \approx \epsilon_e\,E_{\rm
  K,iso}
\end{equation}
For the optical afterglow flux, the synchrotron model predicts that the optical band is below the synchrotron cooling frequency so that:
\begin{equation}
F_{\rm\nu,opt}\propto E_{\rm
  K,iso}^{(p+3)/4}\, n_0^{1/2}\, \epsilon_e^{p-1}\,
\epsilon_B^{(p+1)/4}.
\end{equation}

There are several observational consequences for the two spectral regimes. Firstly, the X-ray luminosity should be independent of $n_0$ and $\epsilon_B$. Secondly, the distributions of the microphysical parameters will introduce scatter in the afterglow luminosity and $E_{\rm K,iso}$ correlation, and in the optical band, $n_0$ and $\epsilon_B$ should introduce additional scatter. For the X-ray band, 
$E_{\rm,iso}$ is a reasonable proxy
for the more directly relevant, but not directly measurable $E_{\rm K,iso}$.

In the context of this work (and other studies) it is important to consider the effect of the Malmquist bias on the $E_{\rm iso}-L_{\rm X}$ correlation using different data selections. For a highly biased sample, where the Malmquist correlation is significant, the intrinsic E$_{\rm iso}$ -- L$_{\rm opt,X}$ correlation can be falsely increased, because both $E_{\rm iso}$ and afterglow luminosity are both determined from flux limited (redshift dependent) observations. The $E_{\rm iso}$ distribution is biased by the sensitivity of {\it Swift}, producing a positive $z-E_{\rm iso}$ correlation. Secondly, the afterglow luminosity distribution is biased by telescope sensitivity. This affect is strongest in the optical (see Table Figure 1.), because redshifts for LGRBs are mostly obtained directly from the optical afterglow (not the host galaxy); this causes a bias for sampling the optically brightest part of the $L_{\rm opt}$ distribution \cite[see][]{2013MNRAS.432.2141C}. Hence, the two independent variables (ignoring the standard synchrotron model for now), $L_{\rm opt, X}$ and $E_{\rm iso}$, are forced to correlate positively via redshift, but possibly only limits for any intrinsic correlations can be inferred \citep{1992ApJ...399..345E}.

\section{Data selection, analysis and results}
\subsection{Data selection}
\subsubsection{Optical, X-ray luminosity, and $E_{\mathrm{iso}}$ data}
LGRB and SRGB optical luminosities, are obtained from K10 and references therein. Our optical data selection taken from K10 includes 61 LGRB optical afterglow absolute magnitudes measured at 24 hr in the rest frame. Based on \citet{norris_2011}, we classify a subset of SGRBs as SGRB-EE in this sample. X-ray luminosities are calculated using light curves from the {\it Swift}-XRT light curve repository \citep{Evans_07,2009MNRAS.397.1177E}. We used an interpolation procedure using flux data in the light curves to find both X-ray luminosities and uncertainties at 11 hr. Burst classifications are based on the scheme from \citep{Howell2014MNRAS} with redshifts taken from the Jochen Greiner online catalogue of localized GRBs\,\footnote{\url{http://www.mpe.mpg.de/~jcg/grbgen.html}}. We note the classification and redshifts in this catalogue are subject to ongoing updates.

To ensure a consistent sample of $E_{\mathrm{iso}}$ data, we use the Butler online catalogue \emph{Swift} BAT Integrated Spectral Parameters\footnote{\url{http://butler.lab.asu.edu/Swift/bat_spec_table.html}}. This catalogue, an extension of \citet{Butler2007ApJ,Butler_2010}, circumvents the nominal BAT upper energy of 150 keV to produce values of $E_{\mathrm{iso}}$ through a Bayesian approach. 
We note that Butler provides no $E_{\mathrm{iso}}$ data for GRBs 061006, 061210 and 071112C, so we use estimates provided by K10.

\begin{figure}
\centering
\includegraphics[scale=0.65]{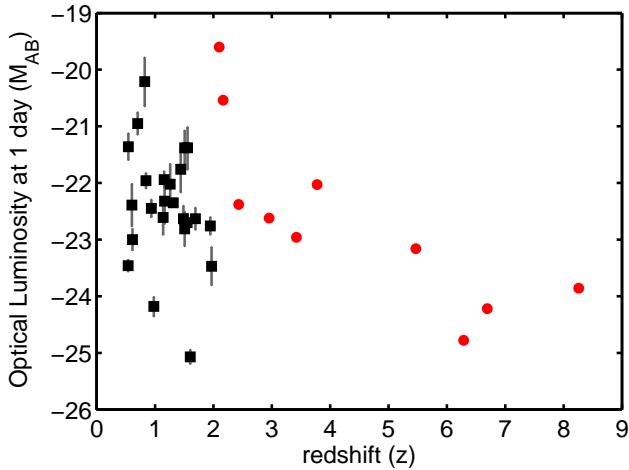}
\caption{Plot of LGRB optical luminosity vs redshift, using a selection from \citet{2010ApJ...720.1513K} (black squares) selected by the following joint selection criteria: $m<24$ and $z<2$, obtained by an iterative procedure that minimizes the correlation (Malmquist bias). Alternatively, the selection $m>24$ and $z>2$ (red circles), is dominated by the Malmquist bias, demonstrated by a significant correlation with a Spearman's probability of a random correlation $p=0.002$. Similarly, but independently, $E_{\rm iso}$ is biased by the sensitivity of {\it Swift}, producing a positive $z-E_{\rm iso}$ correlation (see Table 1).} \label{fig1}
\end{figure}

\begin{figure}
\centering
\includegraphics[scale=0.60]{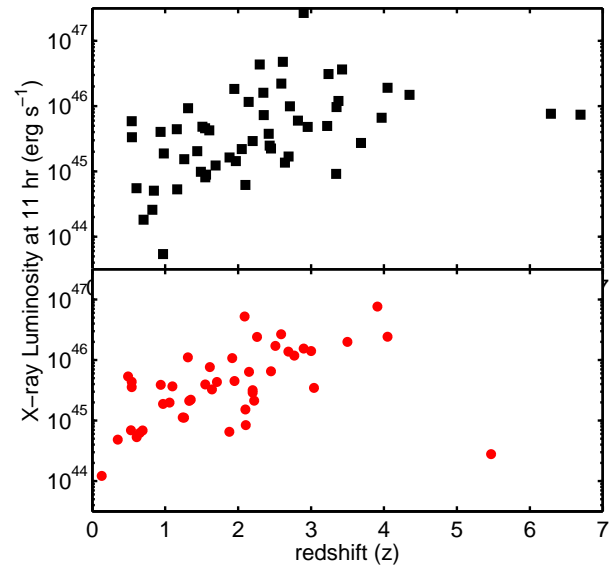}
\caption{ {\bf Top panel} Plot of X-Ray luminosity at 11 hr versus redshift for LGRBs for the optically selected data (an optical afterglow was measurable at 24 hr) taken from Kann et al. (2010). It is clear there is a significant correlation (Spearman $p=10^{-5}$) between X-ray luminosity and redshift, a result of a strong Malmquist bias (flux limited bias), as shown in Figure 1. {\bf Bottom} Same as top panel, but using the  {\it complete} sample from \citet{2009A&A...498..711D}, which was obtained by applying a cut-off in the high energy flux and other observation based selection criteria. Both data selections are equally dominated by the same Malmquist bias (Spearman $p=10^{-5}$).} 
\end{figure} \label{fig2}

\subsection{Analysis and Results}\label{analysis}
\subsubsection{Removing the Malmquist bias}\label{malmresults}
We minimize the Malmquist bias for LGRB optical afterglows using two joint selection criteria: $z<z_{\rm lim}$ and $m<m_{\rm lim}$; i.e. the limiting redshift and apparent magnitude respectively. They are obtained by incrementally reducing the maximum allowable redshift (volume) and magnitude (flux limit) until the correlation between luminosity and redshift is insignificant (i.e. Spearman's $p>0.02$). Using this procedure we obtain 24 samples, the optically unbiased sample. For the optical afterglow luminosity data, the optimal selection criteria that minimizes the bias are: $z<2$ and $m<24$. We also tested a high energy GRB peak flux cut-off, similar to  \cite{2009A&A...498..711D}, and find that the bias in afterglow luminosity vs redshift is not removed. Figure \ref{fig1} plots optical luminosity vs redshift using the Malmquist bias corrected sample, and the highly biased sample: $z>2$ and $m>24$.


Table 1. summarises the correlations and their significance.
After applying the selection criteria above, we find that the correlation between $L_{\rm opt}$ and equivalent isotropic energy $E_{\rm iso}$ in the 24 hr rest frame is significantly weakened and statistically not significant (Spearman $p>0.2$). Alternatively, we find that the E$_{\rm iso}$ -- L$_{\rm opt,X}$ correlation for the data selection $z>2$ and $m>24$ (biased sample) is significant, with Spearman $p=0.002$. 
To test this result for small number statistics (last column Table 1.) we applied a random sampling of 10 events from the full $L_{\rm opt}$ data set, and find that the probability that the observed Spearman's $p<0.002$ is not because of random sampling is about 97\%. This is clear evidence that the E$_{\rm iso}$ -- L$_{\rm opt,X}$ correlations and analysis reported elsewhere in the literature are strongly influenced by selection biases.

For our LGRB X-ray luminosity data (at 11 hr rest frame), we employ the same data selection from the optical sample, namely those X-ray luminosities with an optical afterglow that satisfy $z<2$ and $m<24$ from the \citet{2010ApJ...720.1513K} sample. We tested both the biased X-ray luminosity data (55 LRGBs), and the bias corrected selection (21 LGRBs), for a E$_{\rm iso}$ -- L$_{\rm X}$ correlation. Similar to the optical, we find the biased sample has a significant correlation $(p=10^{-5})$. The same test applied to the bias reduced sample gives a statistically insignificant correlation, similar to that found in the optical sample. 
It is clear that a Malmquist bias is the dominant effect in the observed E$_{\rm iso}$ -- L$_{\rm opt,X}$ correlations in this study, and likely other works. 

\subsection{Synchrotron model comparison to the SGRB and LGRB $L_{\rm opt,X}-E_{\rm iso}$ correlation}
For SGRBs, we followed the same analysis procedure as above to identify a Malmquist bias. Unfortunately, although we identify a distance-dependent bias in SGRB X-ray luminosity, the data is too small (9 bursts with both confident redshifts and X-ray luminosities at 11hr) to apply a robust selection criteria that would yield a statistically significant sample. Hence, we do not apply any selection criteria to this data, but note that the E$_{\rm iso}$ -- L$_{\rm opt,X}$ correlation using this data is likely biased. With this proviso, we find a relatively significant $L_{\rm opt,X}-E_{\rm iso}$ correlation with $p=0.01$.

As shown above, the LGRB data selection (bias free) shows no significant correlation between $L_{\rm opt,X}$ and $E_{\rm iso}$, hence a fit to this data is both unreliable and not meaningful \citep[see][for the pitfalls of fitting to weakly correlated data]{1990ApJ...364..104I}. Alternatively, we can apply a constrained synchrotron model fitted to the median of the LGRB $E_{\rm iso}$ and $L_{\rm opt,X}$ distributions to compare with the SGRB data. We describe the procedure and motivation below:  

 We assume that the microphysical parameter distributions of $\epsilon_B$, $\epsilon_e$ and $n_0$ vary between SGRBs and LGRBs.  We use separate synchrotron models, assuming the same power law index, $p=2.4$, to the SGRB and LGRB data selections, constrained by the median of $E_{\rm iso}$ and $L_{\rm opt,X}$ distributions for the two GRB classes. The result of assuming two different microphysical distributions is to introduce different scalings for $L_{\rm opt,X}$. The ratio of the two scalings is related to differences in the median of the microphysical parameter distributions, and/or energy efficiency, between LGRBs and SGRBs. Finally, the ratio between the optical and X-ray luminosity ratios provides insight into the relative difference in $n_0$ between LGRBs and SGRBs, as the X-ray luminosity is expected to be independent of $n_0$ and $\epsilon_B$.  Finally, we fit a single unconstrained power law model, with the power index a free parameter, for a comparison with other studies. It is not used to place constraints on the microphysical parameter distributions, because of the large scatter of the $E_{\rm iso}$ and $L_{\rm opt,X}$ distributions.

Figure 3 plots $E_{\rm iso}$ and $L_{\rm opt,X}$ for the two scenarios above using the LGRB optical luminosity bias free selection (see Fig 1) combined with SGRB, and SGRB-EE data.
We apply two synchrotron models constrained by the medians of $E_{\rm iso}$ and $L_{\rm opt}$ distributions, assuming $p=2.4$, so that $L_{\rm opt} \propto E_{\rm iso}^{1.1}$ for LGRBs and SGRBs. 
There is $2\sigma$ evidence for different scaling of the two models with ratio of $0.2\pm 0.1$ between LGRBs and SGRBs. 
For X-ray luminosity, the synchrotron model predicts that $L_{\rm X}$ is independent of the microphysical parameters, so that the ratio $L_{\rm X, LGRB}/L_{\rm X, SGRB}$ should be unity. This is shown to be the case in the bottom figure for X-ray luminosity. If we reasonably assume that the scaling is a result of the difference in microphysical parameters between LGRBs and SGRBs, then the scaling ratio implies that the combined microphysical parameter values are relatively larger for SGRBs compared to LGRBs. For comparison with previous studies, we apply an unconstrained power law model fit to the combined LGRB and SGRB data, and find
$L_{opt,24} \sim 1.96 \times 10^{43} E^{0.66}_{\gamma,\mathrm{iso},51}$ erg s$^{-1}$ and $L_{X,11} \sim 2.65 \times 10^{44} E^{0.53}_{\gamma,\mathrm{iso},51}$ erg s$^{-1}$ for the optical and X-ray luminosities respectively. The shallower fits to the combined data, also identified by \cite{2013arXiv1311.2603B}, can be partially explained by the scatter in the two-parameter relation of gamma-ray energy release and X-ray luminosity, which may be attributed to the presence of hidden variables like $E_p$ \citep{2013MNRAS.428..729M}.
\begin{table}
 \begin{tabular}{@{}lccccc}
\hline
\hline
Sample & selection & Correlation & $\rho$ & $p$ & $p_s$(\%)\\
& criteria &  &  &  &\\
\hline
KO$_{61}$ & none  &$z-L_{\rm opt}$ & 0.6 & 10$^{-8}$ & -\\
KX$_{55}$ & none  &$z-L_{\rm X}$ & 0.5 & 10$^{-5}$ & -\\
KB$_{61}$ & none  &$z-E_{\rm iso}$ & 0.4 & 10$^{-3}$ & - \\
\hline
KO$_{61}$  & none  &E$_{\rm iso}$ -- L$_{\rm opt}$ & 0.4 & 10$^{-3}$ & -\\
KX$_{55}$  & none &E$_{\rm iso}$ -- L$_{\rm X}$ & 0.4 & 10$^{-3}$  & -\\
\hline
KOBR$_{24}$ &$m<24, z<2$   &E$_{\rm iso}$ -- L$_{\rm opt}$ & 0.25 & 0.23 & 80 \\
KXBR$_{22}$  &$m<24, z<2$  & E$_{\rm iso}$ -- L$_{\rm X}$ & 0.27 & 0.28  & 80\\
\hline
KOMB$_{10}$ &$m>24, z>2$  &E$_{\rm iso}$ -- L$_{\rm opt}$ & 0.8 & 10$^{-2}$ & 97 \\
\hline
\hline
\end{tabular}
\caption[]{Summary of correlations using different data selections. First column key: KO$_{61}$ = 61 LGRBs optical luminosities at 24 hr rest frame with redshifts taken from K10.  KX$_{55}$ = 51 X-ray luminosities at 11hr matched with KO$_{61}$. KB$_{61}$ is  $E_{\rm iso}$ data from the extended catalogue of \citet{,Butler_2010} and matched to KO$_{61}$. KOBR$_{24}$, KXBR$_{22}$ and KOMB$_{10}$ are LGRB selections satisfying the selection criteria in column 2. Column 3 is the correlation pairs, columns 4 snd 5 are Spearman's $\rho$ and Spearman's $p$ respectively, and column 6 is the probability that the correlations are not a result of small number statistics. See \S\ref{malmresults} for the interpretation of these results.}
\end{table}\label{table1}
\begin{figure}
\centering
\includegraphics[scale=0.62]{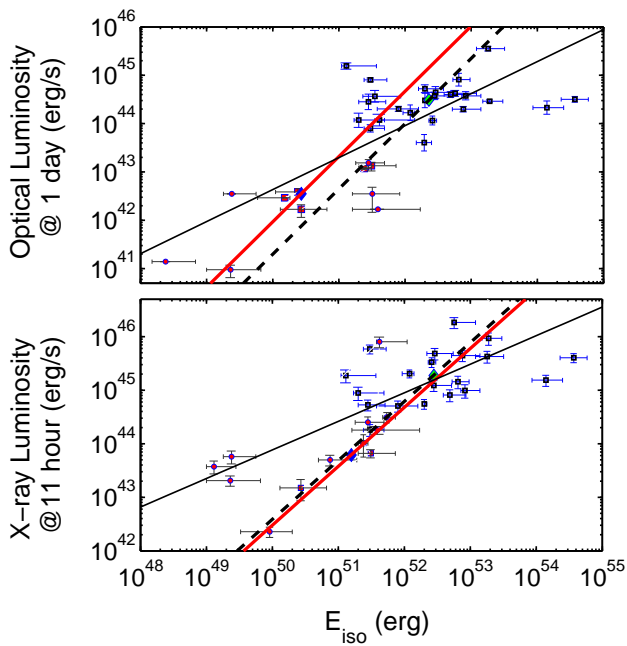}
\caption{ {\bf Top} plot of $E_{\rm iso}$ vs $L_{\rm opt}$, using the bias free selection of LGRB optical luminosity--black squares (see Fig 1), combined with SGRB--magenta circles, and SGRB-EE data--magenta squares. The two lines are synchrotron models for LGRBs (dashed line) and SGRBs (red solid line), constrained by the respective medians of $E_{\rm iso}$ and $L_{\rm opt}$ distributions, assuming $p=2.4$, so that $L_{\rm opt} \propto E_{\rm iso}^{1.1}$. The difference in scale between the two curves, $L_{\rm opt, LGRB}/L_{\rm opt, SGRB}$ is about 0.2. On the same plot, we apply an unconstrained power law model fit $L_{opt,24} \sim 1.96 \times 10^{43} E^{0.66}_{\gamma,\mathrm{iso},51}$ erg s$^{-1}$.  {\bf Bottom} Same as top, but using X-ray luminosity data corresponding to the bias free optical selection above. In contrast to optical luminosity, the synchrotron model predicts that $L_{\rm X}$ is independent of the microphysical parameters $\epsilon_B$ and $n_0$. This is supported by the ratio $L_{\rm X, LGRB}/L_{\rm X, SGRB}\sim1$. The unconstrained fit is $L_{X,11} \sim 2.65 \times 10^{44} E^{0.53}_{\gamma,\mathrm{iso},51}$ erg s$^{-1}$. } 
\end{figure} \label{fig3}

\section{Conclusions}
In summary, we identify a strong Malmquist bias in GRB optical and X-ray afterglow luminosity data. We show that selection effects dominate the observed E$_{\rm iso}$ -- L$_{\rm opt,X}$ correlations. The significance of GRB E$_{\rm iso}$ -- L$_{\rm opt,X}$ correlations depend on the data selection. After removing a Malmquist bias, the E$_{\rm iso}$ -- L$_{\rm opt,X}$ correlation for long GRBs is not statistically significant, but combining SGRB, SGRB-EE and LGRB data, the correlation is significant. It is possible that the correlation between LGRB optical luminosity and redshift is intrinsic (there is a physical correlation between afterglow luminosity and redshift). We do not consider this in our analysis, because it would require including ad-hoc additions to the standard synchrotron model that cannot be currently physically justified. 

Applying a synchrotron model (assuming $p=2.4$) constrained by the median of the E$_{\rm iso}$ -- L$_{\rm opt,X}$ distributions for LGRBs and SGRBs separately, we find a factor of 5 difference in scaling between optical and X-ray luminosities. We test several possible scenarios to explain this result. Firstly, if the prompt energy efficiency for SGRBs was systematically smaller, relative to LGRBs, then the median SGRB $E_{\rm iso}$ would be smaller. But we highlight that the scaling between SGRBs and LGRBs for the X-ray data is unity (fig 3), so this explanation is not consistent with the data. Secondly, if SGRBs had a median beaming angle significantly greater than LGRBs, this would also shift the median SGRB $E_{\rm iso}$ to a smaller value compared to the LGRB $E_{\rm iso}$. But, again this should be apparent in the X-ray $L_{\rm X}-E_{\rm iso}$ correlation. Hence we attribute the difference in scaling between the optical and X-ray correlation as a result of systematic differences in the microphysical parameter distributions of $\epsilon_B$ and $n_0$. Furthermore, for $n_0$ to be similar to (or systematically smaller) for SGRBs compared to LGRBs, requires $\epsilon_B$ to be systematically higher relative to LGRBs.

\vspace{-5mm}
\section*{Acknowledgments}
D.M. Coward is supported by an Australian Research Council Future Fellowship (FT100100345). E.J. Howell acknowledges support from a University of Western Australia Fellowship. This work made use of data supplied by the UK Swift Science Data Centre at the University of Leicester. We thank the reviewer for providing detailed tests and suggestions that have improved the clarity of our results.

\label{lastpage}
\end{document}